# First faint dual-field phase-referenced observations on the Keck interferometer

J. Woillez*[a], P. Wizinowich[a],
R. Akeson[b], M. Colavita[c], J. Eisner[d], R. Millan-Gabet[b], J. Monnier[e], J.-U. Pott[a,f], S. Ragland[a],
E. Appleby[a], A. Cooper[a], C. Felizardo[b], J. Herstein[b], O. Martin[a], D. Medeiros[a], D. Morrison[a],
T. Panteleeva[a], B. Smith[a], K. Summers[a], K. Tsubota[a], C. Tyau[a], E. Wetherell[a]

[a]W. M. Keck Observatory, 65-1120 Mamalahoa Highway, Kamuela, HI 96743, USA
[b]NExScI, California Institute of Technology, 770 South Wilson Ave, Pasadena, CA 91125, USA
[c]JPL, California Institute of Technology, 4800 Oak Grove Drive, Pasadena, CA 91109, USA
[d]University of Arizona, 933 North Cherry Avenue, Tucson, AZ 85721-0065, USA
[e]University of Michigan, 941 Dennison Bldg, AnnArbor, MI 48109-1090, USA
[f]Max Planck Institute for Astronomy, Konigstuhl 17, DE 69117 Heidelberg, Germany

**ABSTRACT**

Ground-based long baseline interferometers have long been limited in sensitivity by the short integration periods imposed by atmospheric turbulence. The first observation fainter than this limit was performed on January 22, 2011 when the Keck Interferometer observed a K=11.5 target, about one magnitude fainter than its K=10.3 limit. This observation was made possible by the Dual Field Phase Referencing instrument of the ASTRA project: simultaneously measuring the real-time effects of the atmosphere on a nearby bright guide star, and correcting for it on the faint target, integration time longer than the turbulence time scale are made possible. As a prelude to this demonstration, we first present the implementation of Dual Field Phase Referencing on the interferometer. We then detail its on-sky performance focusing on the accuracy of the turbulence correction, and on the resulting fringe contrast stability. We conclude with a presentation of early results obtained with Laser Guide Star AO and the interferometer.

## 1 INTRODUCTION

ASTRA, the ASTrometric and phase-Referencing Astronomy upgrade funded by the National Science Foundation, aimed at expanding the capabilities of the Keck interferometer in three incremental steps. First, a self-phase-referencing (SPR) instrument, based on an on-axis fringe tracker, made possible longer integrations for observations with an R~1000 spectral resolution[1]. SPR was intended to benefit an on-going YSO program. Second, a dual-field phase-referencing mode, based on an off-axis fringe tracker, is allowing observation of fainter objects (K<15) with a nearby guide star. This mode will mostly benefit extra-galactic astronomy, increasing by an order of magnitude the number of observable AGN. Third, a narrow angle astrometry mode would measure relative positions with precision of 30 to 100 micro-arcseconds for pairs of objects separated by up to 30 arcseconds. Narrow Angle Astrometry would be used to further characterize known multi-planet systems. Ultimately, combining the upgraded astrometric capabilities of the interferometer with laser guide star adaptive optics on both telescopes would make possible the astrometric monitoring of the inner stars of the galactic center, probing general relativity in the strong field regime. The instrumental and astrophysics goals of the ASTRA project are summarized in Figure 1.

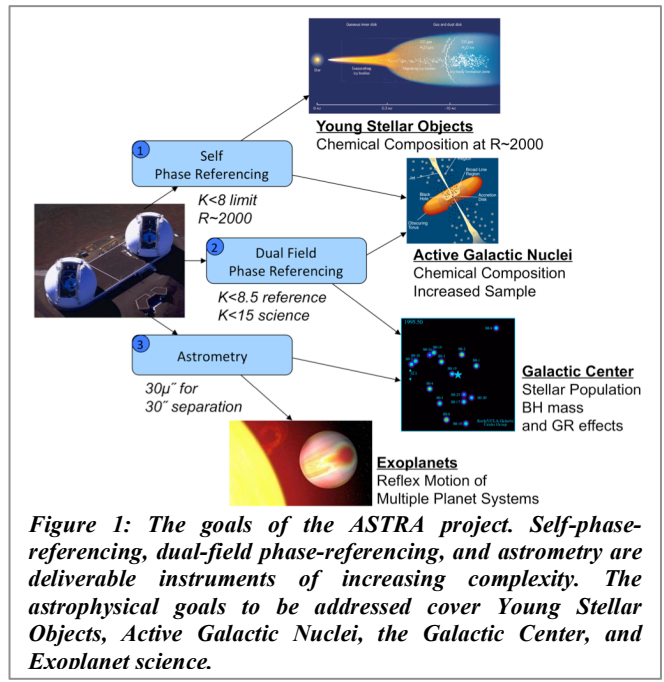

*Figure 1: The goals of the ASTRA project. Self-phase-referencing, dual-field phase-referencing, and astrometry are deliverable instruments of increasing complexity. The astrophysical goals to be addressed cover Young Stellar Objects, Active Galactic Nuclei, the Galactic Center, and Exoplanet science.*

---

* jwoillez@keck.hawaii.edu; phone +1 (808) 881 3511; fax +1 (808) 881 3535; www.keckobservatory.org

Self-Phase-Referencing[1] (SPR) is the first instrument delivered by the ASTRA project to the Keck Interferometer. It has been offered to the Keck community since 2008, and has been used for a wide range of investigations[2][3][4]. NASA's decision to stop funding Keck Interferometer operations means that the 2012A observing semester (February to July, 2012) will be the last scheduled semester for this instrument; all developments related to the Astrometric instrument were therefore put on hold. As such, this contribution focuses on the recent progress and status of the Dual-Field Phase Referencing (DFPR) instrument. Section 2 details the DFPR instrument. Section 3 presents the first on-sky demonstrations and section 0 describes its performance. Section 5 introduces early results combining DFPR and Laser Guide Star Adaptive Optics (LGS-AO).

## 2 DUAL-FIELD PHASE-REFERENCING IMPLEMENTATION

The DFPR instrument is an evolution of the SPR instrument. Instead of using the necessarily bright object of interest to stabilize the fringes, a bright nearby guide star is used for stabilization of a faint object of interest, making long integrations possible. The following sections provide a description of the key components of this instrument: the dual-field subsystem, internal optical path control, internal tip-tilt control, and phase-referencing architecture.

### 2.1 Dual-field subsystem

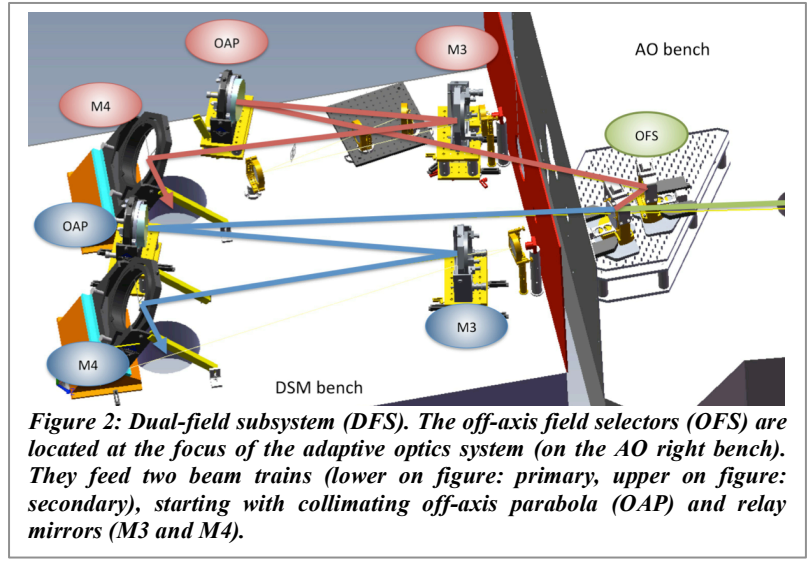

The first element of the dual field subsystem (DFS) shown in Figure 2 is the off-axis field selector (OFS), located on the Adaptive Optics (AO) optical bench at the output focus of the AO system. A hole in the first mirror of the OFS transmits an on-axis star to the primary beam train. The two steerable flat mirrors of the OFS are used to select a star, between 3 and 30 arcsec off-axis, to send through the secondary beam train. After the OFS, both the primary and secondary diverging beams are recollimated with off-axis parabolas, and sent through the primary and secondary beam trains (both already used by the Nuller instrument) to the delay lines and fringe trackers.

*Figure 2: Dual-field subsystem (DFS). The off-axis field selectors (OFS) are located at the focus of the adaptive optics system (on the AO right bench). They feed two beam trains (lower on figure: primary, upper on figure: secondary), starting with collimating off-axis parabola (OAP) and relay mirrors (M3 and M4).*

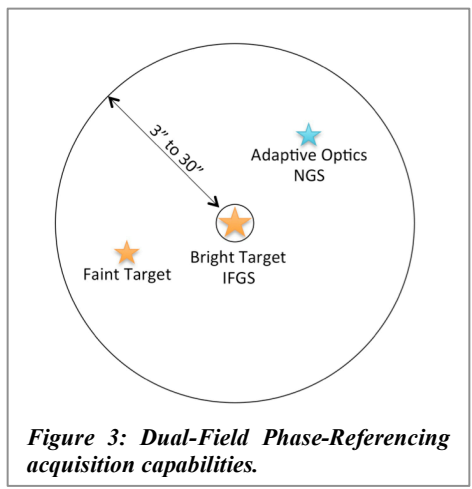

*Figure 3: Dual-Field Phase-Referencing acquisition capabilities.*

The OFS stays fixed throughout an observation. It moves at acquisition time to match the position of the two observed objects within the stable field delivered by the AO system (position angle mode of the AO rotator). The OFS are driven by the interferometer sequencer, computing the position of the off-axis target with respect to the on-axis target, itself positioned at the center of the annular mirror by an appropriate offset of the AO/telescope. At the motion control level, for each new off-axis position requested, the OFS avoids sending the secondary field through the central hole of the annular mirror, therefore preventing longitudinal metrology breaks (unacceptable only for astrometry). From the target acquisition perspective, the interferometer guide star (IFGS) can be selected by an AO system offset pointing up to 30 arcsec away from the AO guide star in Natural Guide Star (NGS) mode, 60 arcsec in Laser Guide Star (LGS) mode; the IFGS can also be the AO guide star. The faint target of interest can be selected by the OFS offset pointing up to 30 arcsec away from the IFGS. Figure 3 illustrates these acquisition capabilities. This operation is completely automated through the interferometer sequencer and can also involve controlling the location of the LGS of the AO system, usually kept on the IFGS for optimal fringe tracking performance.

## 2.2 Internal optical path control

The Keck interferometer is equipped with a set of sensors responsible for minimizing the optical path difference fluctuations introduced by the instrument. For the DFR instrument, they not only minimize the amount of correction the phase-referencing fringe tracker needs to provide (contributing to the instrument ensitivity), but also compensate for differential OPD between the two fringe trackers (making phase-referencing possible). Figure 4 summarizes the various aspects of the internal OPD control. Infrared and internal metrologies are detailed in section 2.2.1, accelerometers in section 2.2.2, and AO Tip-Tilt induced piston compensation in section 2.2.3. The phase-referencing architecture is covered later, in section 2.4.

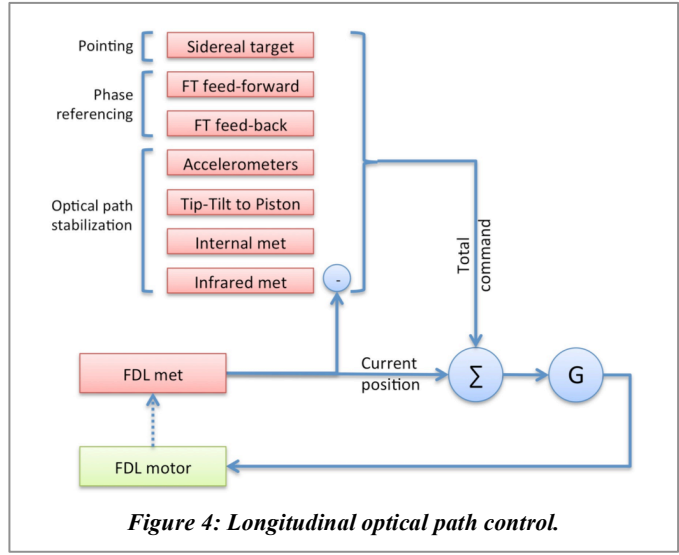

*Figure 4: Longitudinal optical path control.*

### 2.2.1 Longitudinal metrology

The main component of the internal optical path control is an infrared 1319 nm (JDSU NPRO-125N-1319) heterodyne metrology system, named AMET (for Astrometric METtrology). It is very similar in concept to the metrology gauge[6] designed for the Differential Phase instrument[7], where orthogonal masks are used to isolate two metrology propagations with a low level of self-interference and cross-talk. In the ASTRA implementation, an annular mask (conveniently letting the star light through) selects the propagation through the beam train up to a corner cube located at a pupil location inside each AO system, while a central mask selects the annular propagation to a reference corner cube located on the opposite side of the injection beam splitter. Each of the four metrologies, one per beam train, will measure the difference between the optical path from the injection dichroic to the metrology corner cube, and the optical path from the dichroic and the reference corner cube. To complete the beam train coverage, another existing metrology system (named Internal) based on a single stabilized HeNe laser source, measures the on-axis optical path difference from the beam combiner beam splitters to the reference corner cubes. The propagation selection for this visible system is based on selecting orthogonal polarizations with polarizers. For the reference corner cube to work with both systems, it is composed of an on-axis inner mask that blocks the infrared system, and also selects one single linear polarization for the visible system. Figure 5 gives a summary of the AMET and Internal setup. The detection electronics for the infrared system is an exact replica of the 633 nm metrology systems, except for the detector cards modified to operate at 1319 nm. It is capable of measuring optical paths evolution with an accuracy of 2.5 nm.

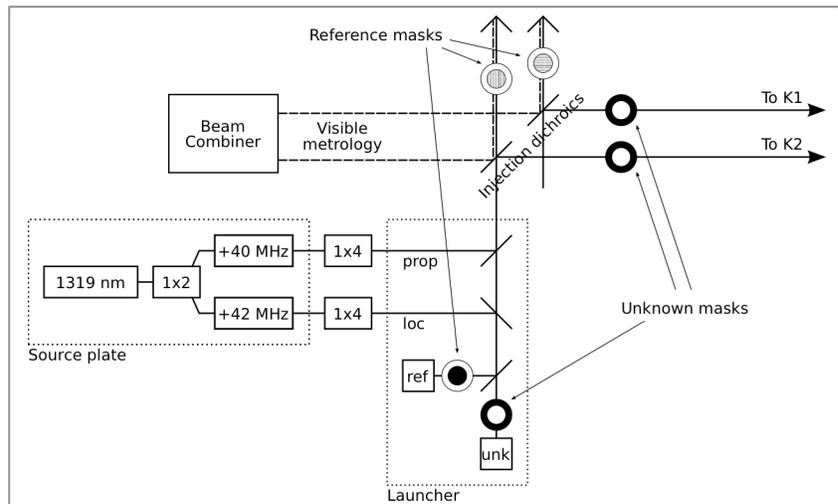

*Figure 5: Longitudinal metrology layout. The infrared reference output measures the optical path from the launcher to the reference corner cube, the infrared unknown output measures from the launcher to the metrology retro-reflector (not shown) inside the AO systems, and the visible metrology measures in differential from beam combiner beam splitter to the reference corner cubes. All three combined, the longitudinal metrologies measure the OPD from the beam combiner dichroics to the common metrology corner cubes in the AO systems, providing complete differential beamtrain coverage.*

The single infrared laser source of the AMET does not have a stabilized wavelength. Instead, we have chosen to operate it at a differential optical path difference (OPD) close to zero, in order to cancel any impact of the wavelength fluctuation on the differential OPD control. This can be achieved by adjusting the position of the reference corner cubes to match the differential OPD in the delay lines. In practice, this is equivalent to a transfer of the differential OPD measurement to the visible HeNe metrology system, which already meets the wavelength stability requirement, without need of an additional stabilization. However, the wavelength fluctuations can cause a degradation of the OPD control of a single beam train, but on the fast fringe tracker timescale of typically 1 ms, the amplitude of the effect is estimated to be a negligible 2 nm for a maximum optical path difference of 85 m (laser line width < 5 kHz/ms). On second timescales, this wavelength fluctuation appears as an additional turbulence-like term of 37 nm rms over the same 85 m delay (estimated from a laser frequency drift < 50 MHz/hour), very well correlated between primary and secondary targets, and therefore perfectly well corrected by the bright primary fringe tracker.

Both visible and infrared systems are relative metrology systems: they only provide an accurate optical path variation relative to an offset, which can be assumed fixed as long as the metrology systems stay uninterrupted. The determination of this offset, only needed for astrometry, would have been achieved through a swap of the observed pair of targets between the primary and secondary beams. This operation corresponds to a sign change for the separation vector, which makes the half sum of differential OPD give directly the metrology offset. This swap operation would have been the basis of an astrometric measurement for the astrometric implementation.

Finally, each fast delay line having a local metrology system, the difference between longitudinal metrology and delay line local metrology can be sent for compensation to the delay line servos (see Figure 4).

### 2.2.2 Accelerometers

The same accelerometers used by other Keck Interferometer instruments participate to the reduction of vibrations introduced by the telescopes. But since the longitudinal metrology reaches further inside the AO system, the accelerometer measuring the relative vibration between the AO bench and the dual star module are turned off, while keeping the telescope primary mirror, tertiary mirror, and AO bench accelerometers on. Even though, the secondary mirror is also equipped with accelerometers, their contribution has always been turned off, due to the fact that the focus offload from the AO system to the secondary mirror exceeds the accelerometer dynamic range and causes transient saturation of the accelerometer compensation. Some of the vibrations residuals seen by the fast fringe tracker are certainly coming form the telescope secondary mirrors. The combined accelerometer measurements at each telescope are then sent for compensation to the appropriate delay lines (see Figure 4).

### 2.2.3 AO Tip-Tilt induced piston compensation

The Tip-Tilt mirror of the Keck AO system is not in a pupil plane. Consequently, the positions of the primary and secondary beams are not the same on this mirror, and the mirror motion therefore introduces a variable differential OPD. Using the same infrastructure as the accelerometers, the motion of the Tip-Tilt mirror is digitized and sent to the FDL controllers, where it is projected on the separation between secondary and primary targets. The result is used as an additional target contribution to the FDL servo, as shown in Figure 4. Based on the geometry of the Tip-Tilt mirror inside the AO system, the amplitude of the effect has been estimated to be 523 nm/arcsec$^2$ of induced piston per arcsec of correction and arcsec of separation. For seeing of 0.5 arcsec, and a separation of 15 arcsec, this amounts to ~1 micron differential OPD, or a drastic 95% fringe contrast reduction per telescope on long exposures at 2.2 microns.

## 2.3 Internal Tip-Tilt control

For bright objects, there are usually enough photons for an Angle Tracker to correct and stabilize the pointing of the interferometer beamtrains. However for DFPR, where one of the two objects is faint, alternative method must be used.

### 2.3.1 Tip-Tilt metrology

An internal Tip-Tilt metrology was initially planned for DFPR. It consisted in a 690 nm laser beacon injected in the four beamtrains after the OFS, and 2D position sensitive devices (PSD) located in the basement next to the infrared Angle

Trackers (see Section 2.3.2). The PSDs were supposed to measure any Tip-Tilt introduced by the beam trains after the AO correction, and send a fast correction to the existing Tip-Tilt mirror in the interferometric basement. Such a system would have preserved the Tip-Tilt stability of the AO correction for the interferometric fringe trackers. In addition to the AO residuals, only the Tip-Tilt anisoplanetism would be left uncorrected. However, the low transmission at 690 nm to the PSDs required operating the beacons at a power level high enough to contaminate the AO wavefront sensors; this component was therefore never successfully used on sky. An easy fix, which has not been implemented, would be to relocate the sensors before the fringe trackers pick-up dichroics, which are responsible for most of the losses, either as part of the AMET dichroic, or as dedicated 690 nm pickup dichroics.

### 2.3.2 Blind acquisitions

Without Tip-Tilt metrology, a target too faint to be seen by the Angle Trackers cannot be used to align the beam trains (otherwise the standard acquisition procedure). We are however able to send the alignment laser up from the beam combiners toward the longitudinal metrology retro-reflector in front of the AO deformable mirror. The reflected light is visible on the AO acquisition camera and marks the current pointing direction of the beam trains. This information can then be used to correct the beam train pointing direction to match the separation of the observed object. In a way, this replaces the Tip-Tilt metrology described above, at least at acquisition time.

### 2.3.3 Angle Tracker

The H-band angle tracker, traditionally running at 80 Hz on bright objects, has to be slowed down to reach the magnitude of the faint object found in one of the two beam trains. The achieved correction bandwidth can then only address very slow drifts of thermal or alignment origin, leaving tip/tilt vibrations and beam train turbulence uncorrected. Ideally, one would want a dedicated slow camera for the faint object and keep a fast rate on the bright object. An intermediate solution was implemented with the addition of a variable length average filter on the image data coming from the camera. Increasing the length of the filter is a less readout noise efficient equivalent to slowing down the camera. This approach keeps a good Tip-Tilt control bandwidth on the bright object, where the average filter is disabled.

An optimal approach, requiring only one camera, would have been to remove the destructive reads for the regions of interests containing the faint object. These improvements to the Angle Tracker could have been made in parallel with a functional Tip-Tilt metrology system.

## 2.4 Phase-referencing architecture

Once the instrumental piston and tip/tilt are (almost) properly corrected, the phase-referencing performance relies on the fast on-axis fringe tracker to stabilize the fringes and send the proper correction to the faint off-axis instrument where long integrations are performed.

### 2.4.1 Fringe trackers

The principle and performance of the KI fringe trackers have been documented elsewhere[5]. They are routinely used to perform $V^2$ measurements, as well as to stabilize the OPD for the Nuller instrument. Specific to ASTRA, and already detailed for the SPR instrument[1], two distinct cameras (the fast and slow fringe trackers in Figure 6) are available to operate the two fringe trackers at different rates.

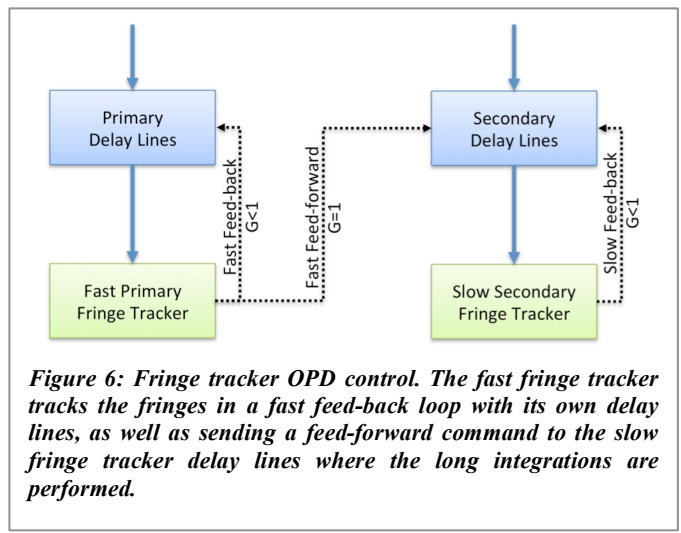

*Figure 6: Fringe tracker OPD control. The fast fringe tracker tracks the fringes in a fast feed-back loop with its own delay lines, as well as sending a feed-forward command to the slow fringe tracker delay lines where the long integrations are performed.*

### 2.4.2 Feed-back and feed-forward

The fast primary fringe tracker operates in closed loop (loop gain lower than 1 for stability reasons) with its own primary delay lines. Since each fringe tracker has its own delay lines, as illustrated in Figure 6, the fast primary fringe tracker is able to send an open-loop correction to the secondary delay lines (loop gain of 1), providing an improved OPD correction for the secondary side. This feed-back/feed-forward architecture principle for the DFPR instrument is similar to SPR. The only major addition is the capability, controlled by the interferometer sequencer, to exchange which of the two fringe tracker is phase-referencing the other, based on the magnitudes of the observed pair (the faint target can be observed on the primary or secondary side).

## 3 DEMONSTRATION

### 3.1 First light: 2MASS J04531720-0755503, K=11.5

On January 22, 2011, the interferometer, configured in DFPR mode, locked its reference fringe tracker running at 4 millisecond integration period on a $K = 7$ mag star and provided a correction for a $K = 11.45$ mag star, 10 arcseconds away. The faint fringe tracker, using a 0.1 second integration time, was able to follow fringes on the faint star and measure a fringe contrast of 0.89 (see Figure 7). An extrapolation, purely based on the fringe tracking performance, gives a limit of $K \sim 15$ mag for the faint objects, to be considered as an upper limit.

### 3.2 Current demonstrated performance: K=12.5

So far, the faintest observed objects have magnitudes of K=12.5. Contrast measurements are also routinely performed following the standard $V^2$ sequence, made of fringe, ratios, and dark measurements[1]. Contrast stability and accuracy have not been fully qualified, and absolute $V^2$ measurement not performed, as described in the following performance section.

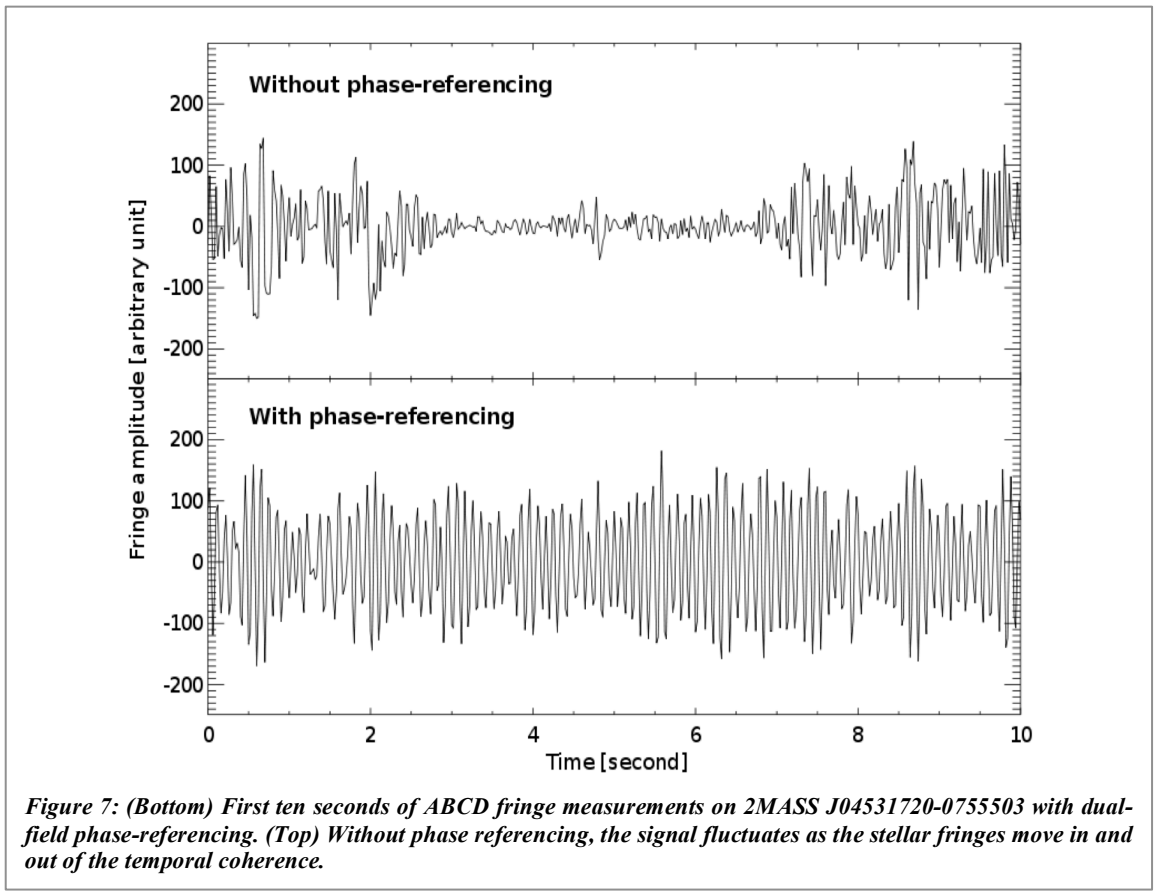

*Figure 7: (Bottom) First ten seconds of ABCD fringe measurements on 2MASS J04531720-0755503 with dual-field phase-referencing. (Top) Without phase referencing, the signal fluctuates as the stellar fringes move in and out of the temporal coherence.*

# 4 PERFORMANCE

To characterize the on-sky performance of DFPR, we have typically performed two kinds of observations. First, detailed in Section 4.1, we present the performance of a single bright fringe tracker. Then, detailed in Section 4.2, we looked into the correlations in OPD measurements between bright pairs of various separations. This shows how well the metrologies correct for instrumental effects, and how close the differential OPD gets to the fundamental limit set by the atmospheric differential turbulence. Last, detailed in Section 4.3, we performed contrast measurements with the slow fringe tracker on the faint object of a Bright-Faint pair. This is exactly the kind of measurement DPFR is designed to perform.

## 4.1 Bright fringe tracker performance

The instrumental vibrations not properly corrected by the internal optical path control infrastructure (accelerometers and metrologies) combined to the atmospheric piston have to be measured and compensated by the phase-referencing fringe tracker. Typical fringe tracker compensations are presented in Figure 8, and highlight the benefits of DFPR. At high frequencies, residual vibrations dominate. They require that the phase-referencing fringe tracker be run at high rate (typically ~250 Hz fringe rate). A better vibration control (working accelerometers on the secondary mirror being the obvious gap) would certainly help slow down the fringe tracker and increase the bright star limiting magnitude for the DFPR instrument, currently estimated at K~8 (this is the $V^2$ limiting magnitude at 250 Hz, reduced by a one magnitude margin for stable fringe tracking performance).

## 4.2 Phase-referencing on bright pairs

By observing bright pairs with DFPR, we are able to measure the amount of decorrelation, introduced by the instrument, between the primary and secondary beam trains. The amount of ΔOPD is computed as the sum of the contributions from the longitudinal metrologies in differential between the primary and the secondary, the differential

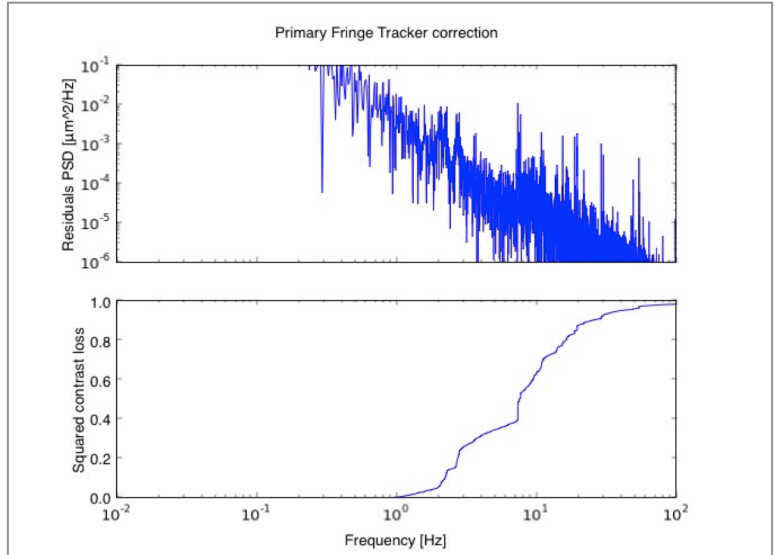

*Figure 8: OPD residuals measured by the bright fast fringe tracker. This is the sum of all the instrumental OPD fluctuations not corrected by the metrologies. Top: Power spectrum density of the OPD fluctuations. Bottom: Hypothetical squared contrast loss if the fringe tracker were slowed down. The contrast loss is already significant at 10 Hz. Most of the degradations come from narrow vibration lines.*

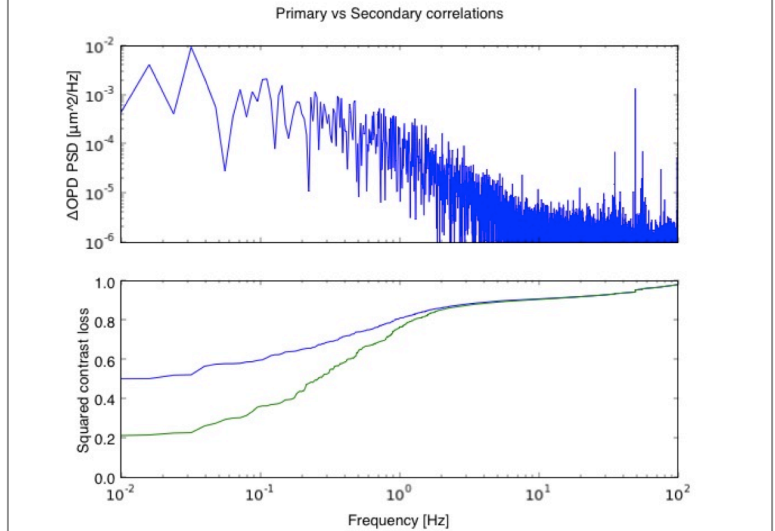

*Figure 9: OPD correlations between primary and secondary beamtrains. – Top: Power Spectrum Density (PSD) of the differential OPD, estimated from the combination of metrologies, fringe trackers phase, and AO Tip-Tilt piston correction. – Bottom: Contrast loss estimated from PSD above versus integration frequency. The upper curve is with AO Tip-Tilt piston correction, the lower curve without. (Bright-Bright pair: 2MASS J01174326-1220413, K=7.5; 2MASS J01174320-1220339, K=7.9; separation = 3.7 arcsec; MJD=55844.45336)*

piston effect from the AO Tip-Tilt, and the fringe tracker phases in differential. Note that the delay in the phase estimator of the Fringe Trackers of 2.4 ms (~3 servo cycles at 1250 Hz servo rate) has been properly compensated. An illustration of our performance is given in Figure 9. This shows how well (165 nm rms) the differential OPD measurements is performed, and gives an upper bound to the $V^2$ on the slow phase-referenced side, that could be achieved if the fast fringe tracker corrections were instantaneous. Very obvious in Figure 10, not compensating the differential piston introduced by the AO Tip-Tilt mirror would cause a noticeable degradation of the fringe contrast for sub-Hz frequencies (already at this small 3.7 arcsec separation).

When phase-referencing is active, the combination of the correction applied and the residual phase error measured by the phase-referenced fringe tracker, gives a direct estimation of the visibility reduction that would impact long integrations. This approach takes into account the limited performance of the control loops (mostly time delays associated to the fast Fringe Tracker correction). Results on a bright pair are presented in Figure 10. The expected squared contrast reduction, for an integration period of 1 second, is ~0.7. This figure also shows that the contrast drops sharply for integration periods above 8 seconds. This has been traced to a high-pass filter ($\tau = 2$ s) applied to the infrared longitudinal metrology (section 2.2.1), needed to prevent rare metrology break events from injecting large position offsets to the tracking delay lines. Longer integrations would require that a larger time constant be used for this high-pass filter.

Even though this was our intent, we are not able to present a systematic study of the evolution of differential OPD as a function of pair separation. This is something that would have required more engineering time than available. More recent attempts to measure correlations for wide separation have been heavily impacted by adverse atmospheric conditions.

### 4.3 Contrast measurements on faint objects

The first contrast measurement with the DFPR instrument was performed in February 2012, on a K=12.4 target with a K=7.9 reference, 14 arcsec away. The results are illustrated in Figure 11. The achieved squared contrast of ~0.2, for this unresolved target, is below the expected 0.56 (0.8 instrumental squared contrast times 0.7 squared visibility loss from Figure 10) we should have achieved for a narrow separation. A strong (potentially high altitude) seeing is certainly behind this additional drop in contrast, and compatible with a poor off-axis AO correction also seen on that night.

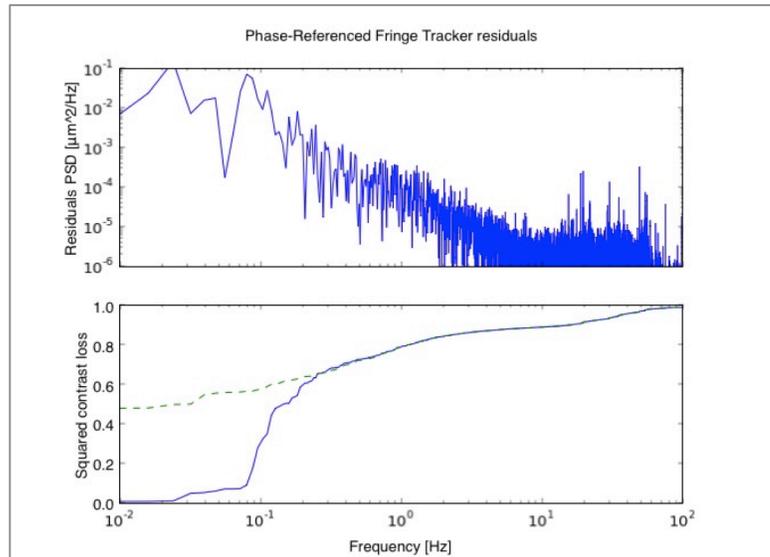

*Figure 10: Residuals measured by the phase-referenced fringe tracker. Top: Residual Power Spectral Density. Bottom: Squared contrast loss. The solid curve is obtained with a high pass filter ($\tau=2s$) on the infrared metrology, the current operation setup; the dashed curve corrects for the missing low-pass component. (Same Bright-Bright pair as in Figure 9.)*

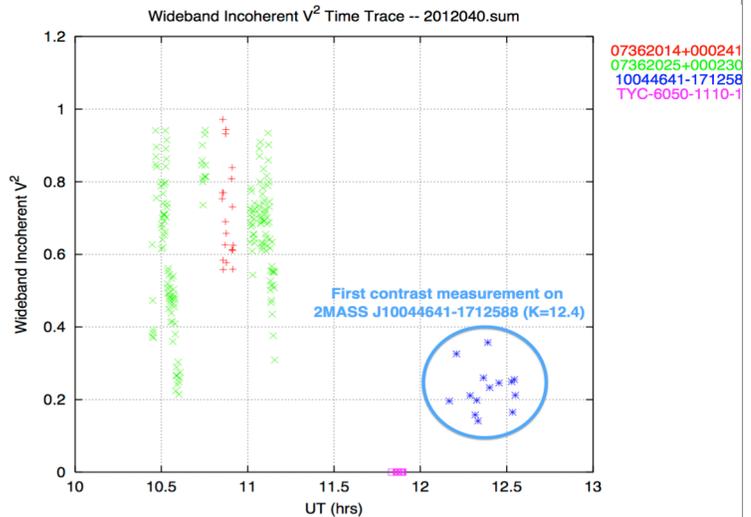

*Figure 11: First squared contrast (~0.2) measurement on 2MASS J10044641-1712588 (K=12.4), with a 14 arcsec guide star TYC-6050-1110-1 (K=7.9). MJD=55966.43403. It was a poor seeing (no MASS-DIMM measurement, 1.2~1.6 arcsec from WRF model) and high wind night (>55 km/h). Note: The square contrast measurements on the bright targets between 10 UT and 11.5 UT were made with the secondary fringe tracker in phase-tracking mode; the drifts come from the phase track point moving away from the maximum of the fringe envelope.*

The next logical step would be to study the temporal stability of the contrast measurements, in order to estimate the accuracy of an absolute $V^2$ measurement with the DFPR instrument. This would require observing and cross-calibrating multiple calibrators. Already, if the absolute $V^2$ stability turns out to be too difficult to estimate, we should be able to perform differential measurement on the faint target in a manner similar to the SPR instrument[1].

## 5    ADAPTIVE OPTICS LASER GUIDE STAR

With a Laser Guide Star being commissioned on the Keck I telescope[8], the Keck Interferometer should benefit from a much-improved sensitivity on very red objects. This was identified by the ASTRA project as a key instrumental development benefiting its Active Galactic Nuclei and Galactic Center programs.

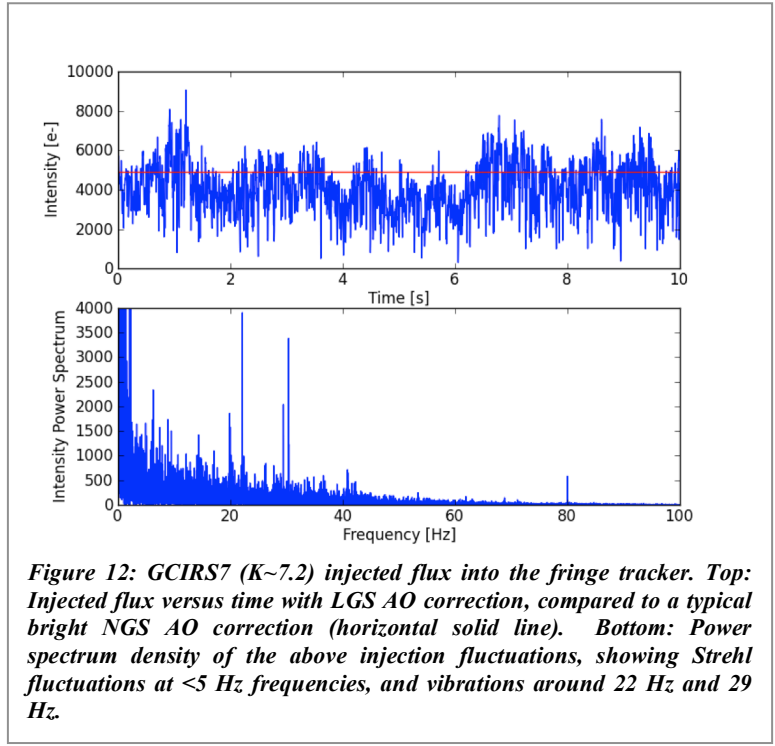

Figure 12: GCIRS7 (K~7.2) injected flux into the fringe tracker. Top: Injected flux versus time with LGS AO correction, compared to a typical bright NGS AO correction (horizontal solid line). Bottom: Power spectrum density of the above injection fluctuations, showing Strehl fluctuations at <5 Hz frequencies, and vibrations around 22 Hz and 29 Hz.

The current limit for on-axis AGN observation is the performance of the NGS-AO correction. The low delivered Strehl impacts both the Fringe Tracker and the Angle Tracker. LGS-AO should make $V^2$ measurements possible on fainter AGN, and even make the SPR instrument usable on the brightest ones (e.g. NGC 4151).

For the Galactic Center, GCIRS7 is bright enough in the K band to be used as the reference for the DFPR instrument. However, previous observation attempts have shown that NGS-AO does not deliver a Strehl high and stable enough (IRS7 itself is too faint for NGS-AO, and the AO guide star is 37 arcsec away) to use it as a phase reference.

For the first time, on April 30, 2012, the LGS-AO corrected Keck II telescope[9] delivered GCIRS7 light to the interferometer fringe tracker. The injected flux level and stability was shown to be comparable to what NGS-AO typically delivers for an R~7 and K~7 target (see Figure 12). An LGS-AO facility has just been commissioned on Keck I. We expect to use both LGS-AO systems this summer to potentially observe Active Galactic Nuclei and the Galactic Center.

## 6    CONCLUSION

Dual field phase referencing has been successfully demonstrated for the first time on the sky and with a significant increase in limiting magnitude. This opens up the possibility to study a much wider range of astrophysical targets at the high spatial resolution offered by interferometry. This work should not yet have a conclusion, but alas it will with the cessation of Keck Interferometer operations at the end of July 2012.

## ACKNOWLEDGMENTS

The W. M. Keck Observatory is operated as a scientific partnership among the California Institute of Technology, the University of California, and the National Aeronautics and Space Administration. The Observatory was made possible by the generous financial support of the W. M. Keck Foundation.

The data presented herein were obtained at the W. M. Keck Observatory, which is operated as a scientific partnership among the California Institute of Technology, the University of California, and the National Aeronautics and Space Administration. The Observatory was made possible by the generous financial support of the W. M. Keck Foundation.

The authors wish to recognize and acknowledge the very significant cultural role and reverence that the summit of Mauna Kea has always had within the indigenous Hawaiian community. We are most fortunate to have the opportunity to conduct observations from this mountain.

This material is based upon work supported by the National Science Foundation under Grant No. AST-0619965. This material is based in part upon work performed for the Jet Propulsion Laboratory, California Institute of Technology, sponsored by the U.S. Government under Prime Contract NAS7-03001 between the California Institute of Technology and NASA.